\documentclass[12pt]{article}

\usepackage[flushleft]{threeparttable}
\usepackage[T1]{fontenc}
\usepackage{amsmath,amssymb}
\usepackage{graphicx}
\usepackage{caption}
\usepackage{color}
\usepackage{dcolumn}
\usepackage{bm}
\usepackage[numbers,super,comma,sort&compress]{natbib}
\usepackage{threeparttable}
\usepackage{multirow}
\usepackage{mhchem}
\usepackage{float}
\usepackage{placeins}
\usepackage{adjustbox}
\usepackage{array,multirow}
\definecolor{background-color}{gray}{0.98}

\title{Modeling the electronic structures of the ground and excited states of the ytterbium atom and the ytterbium dimer: A modern quantum chemistry perspective.}
\author{
Pawe{\l} Tecmer\thanks{Institute of Physics, Faculty of Physics, Astronomy and Informatics, Nicolaus Copernicus University in Torun, Grudziadzka 5, 87-100 Torun, Poland; email: ptecmer@fizyka.umk.pl}, 
Katharina Boguslawski\thanks{Department of Chemistry, Nicolaus Copernicus University in Torun, Gagarina 7, 87-100 Torun, Poland}~\footnotemark[1], 
Mateusz Borkowski\footnotemark[1],\\
Piotr Szymon \.Zuchowski\footnotemark[1],
and Dariusz K\k{e}dziera\footnotemark[2] 
}
\begin{document}

\maketitle

\begin{abstract}
We present a comprehensive theoretical study of the electronic structures of the Yb atom and the Yb$_2$ molecule, respectively, focusing on their ground and lowest-lying electronically excited states.  
Our study includes various state-of-the-art quantum chemistry methods such as CCSD, CCSD(T), CASPT2 (including spin--orbit coupling), and EOM-CCSD as well as some recently developed pCCD-based approaches and their extensions to target excited states.  
Specifically, we scan the lowest-lying potential energy surfaces of the \ce{Yb2} dimer and provide a reliable benchmark set of spectroscopic parameters including optimal bond lengths, vibrational frequencies, potential energy depths, and adiabatic excitation energies.  
Our in-depth analysis unravels the complex nature of the electronic spectrum of \ce{Yb2}, which is difficult to model accurately by any conventional quantum chemistry method. 
Finally, we scrutinize the bi-excited character of the first $^1\Sigma_g^+$ excited state and its evolution along the potential energy surface.  
\end{abstract}

\clearpage


  \makeatletter
  \renewcommand\@biblabel[1]{#1.}
  \makeatother

\renewcommand{\baselinestretch}{1.5}
\newcommand{\down}{\sout{$\downarrow$}}
\newcommand{\up}{\sout{$\uparrow$}}
\newcommand{\zero}{\sout{\phantom{$\downarrow \negthickspace \uparrow$}}}
\newcommand{\double}{\sout{$\downarrow \negthickspace \uparrow$}}
\newcommand{\ab}{\bar{a}}
\newcommand{\ib}{\bar{i}}
\newcommand{\Lower}[1]{\smash{\lower 1.5ex \hbox{#1}}}

\newcommand*{\pawel}[1]{{\textcolor{blue}{PT: #1}}}
\newcommand*{\kasia}[1]{{\textcolor{orange}{KB: #1}}}
\newcommand*{\pzuch}[1]{{\textcolor{red}{PZ: #1}}}
\normalsize

\clearpage
\section*{\sffamily \Large INTRODUCTION} 
The divalent ytterbium atom has in recent years garnered significant attention thanks to its many uses in cold atom physics. 
It has a non-magnetic $^1$S$_0$ ground state and several useful optical transitions: the strong $^1$S$_0\leftrightarrow^1$P$_1$ line can be used for Zeeman slowing, whereas the narrow (181 kHz) intercombination $^1$S$_0\leftrightarrow^3$P$_1$ line can be used to directly laser cool Yb atoms to microkelvin temperatures~\cite{Honda2002}. 
Yb has seven stable isotopes: two fermions (171 and 173 with nuclear spins of $1/2$ and $5/2$, respectively) and five bosons (168, 170, 172, 174, and 176) that lack nuclear spin. 
The rich isotope structure makes it possible to mass-tune the atomic interactions~\cite{Kitagawa2008} and facilitates a wide array of possible quantum-degenerate gases~\cite{Takasu2003,Fukuhara2007,Fukuhara2009,Sugawa2011}. 
The doubly forbidden $^1$S$_0$$\leftrightarrow$$^3$P$_0$ transition lies at the heart of optical atomic clocks~\cite{Ludlow2015} which are among the most precise physical instruments known to mankind. 
For example, an ytterbium clock has recently been demonstrated to enable geopotential measurements with an accuracy below a centimetre~\cite{McGrew2018}. The long lived $^3$P$_0$ clock states also find use in quantum simulations using Yb atoms~\cite{Livi2016}.
	
The long range interactions in the Yb dimer have been probed extensively by high resolution photoassociation spectroscopy (PAS)~\cite{Jones2006} near the narrow $^1$S$_0$$\leftrightarrow$$^3$P$_1$ intercombination line. 
The excited $^1$S$_0$+$^3$P$_1$ ($0_u^+$)~\cite{Tojo2006, Borkowski2009} state has been probed by single color PAS and provided the van der Waals $C_6$ coefficient and an improved value of the atomic $^3$P$_1$ lifetime. 
Two-color PAS of ground state $0_g^+$ vibrational levels~\cite{Kitagawa2008,Borkowski2017a} delivered accurate information about the cold scattering properties of Yb, most notably the $s$-wave scattering lengths for all isotopic combinations. 
Additionally, intercombination line PAS has provided insight into such exotic physical phenomena as subradiant $1_g$ states~\cite{Takasu2012} and hyperfine-induced purely-long-range states~\cite{Enomoto2008a}. 
PA lines near the intercombination line also gives experimentalists the capacity to alter the scattering properties of ultracold Yb atoms optically through the optical Feshbach resonance mechanism~\cite{Ciurylo2005,Enomoto2008,Borkowski2009,Kim2016}.
	
Other excited states of Yb$_2$ have also been probed in ultracold atom experiments. 
The $s$-wave scattering length in the $^1$S$_0$+$^3$P$_0$ ($0_u^-$) state in $^{174}$Yb$_2$ has been determined through clock spectroscopy of Bose--Einstein condensates trapped in 3D optical lattices~\cite{Franchi2017,Bouganne2017}. 
Based on these measurements the positions of near-threshold molecular clock states could be determined~\cite{Borkowski2018}. 
Orbital Feshbach resonances have been utilized to produce strong correlations~\cite{Pagano2015} in a degenerate Fermi gas of $^{173}$Yb atoms and create a novel type of Feshbach molecules~\cite{Cappellini2018}. 
Collisions of atoms in the ground and metastable $^3$P$_2$ states have also been considered: magnetic Feshbach resonances~\cite{Kato2013} and Feshbach molecules~\cite{Takasu2017} have been observed experimentally. 
Feshbach resonances in this pair of states have been shown to exhibit quantum chaos~\cite{Green2016}. 
	
Finally, Yb offers excellent opportunities for tests of fundamental physics. 
It -- uniquely -- possesses two independent clock transitions with an unusually high sensitivity to possible variations of the fine structure constant~\cite{Safronova2018}. 
Open-shell molecules involving Yb, like RbYb or LiYb can be used in searches for the electric dipole moment of the electron (eEDM)~\cite{Meyer2009}. 
Thanks to its simple structure, the ground state Yb dimer is an excellent testing ground for the search for new short scale gravitylike forces~\cite{Borkowski2018a}, temporal variations of the proton-to-electron mass ratio~\cite{Zelevinsky2008}, or beyond-Born-Oppenheimer effects~\cite{Borkowski2017a}. An ``optical molecular clock''~ \cite{Borkowski2018} utilizing forbidden $J=0 \leftrightarrow 0$ transitions between the ground $(1)\,0_g^+$ and clock $(1)\,0_u^-$ states in Yb$_2$ could provide energy level measurements at an accuracy so far unprecedented in molecular spectroscopy.
	
Despite these advancements, our knowledge of the ground and excited state electronic structures of the Yb$_2$ dimer and the resulting interaction potentials remain scarce. 
The detailed knowledge of molecular potential energy curves would be the first step towards the production of deeply bound ultracold molecules via stimulated Raman adiabatic passage (StiRAP)~\cite{Takekoshi2014}. 
Unfortunately, the reliable quantum chemical modeling of the Yb$_2$ dimer is not straightforward and poses a remarkable challenge for present-day quantum chemistry to provide accurate interaction potentials for both the ground and excited states.
The large number of correlated electrons combined with a sizable all-electron basis set makes the theoretical modeling of \ce{Yb2} computationally very demanding.  
Moreover, the Yb atom (\ce{Z=70}) falls into the class of heavy elements and thus requires a relativistic description of the electronic motion. 
While for a qualitative study it is sufficient to account for scalar relativistic effects only, spin--orbit coupling has to be included in calculations for a quantitative analysis as well as for electron excitation energies. 
The rather complex interplay between electron correlation and relativistic effects~\cite{Timo_overview} impedes routine quantum chemical calculations on the \ce{Yb2} dimer. 

To date, the spin--free coupled cluster ground state interaction potential was investigated by Buchachenko~\emph{et al.}~\cite{Buchachenko2007}, while reliable van der Waals \ce{C_6} coefficients for ground and selected excited states were calculated by Safronova~\emph{et al.}~\cite{Safronova2012} and Porsev~\emph{et al.}~\cite{Porsev2014}, respectively. 
The very first attempt to understand and model the low-lying part of the electronic spectrum of Yb$_2$ has been already performed in 1998 by Wang and Dolg~\cite{dolg-yb2}. 
In their multi-reference configuration interaction study, they considered potential energy curves resulting from electron excitations from the occupied $\sigma_u$ and $\sigma_g$ to the virtual $\pi_u$, $\pi_g$, $\sigma_g$, and $\sigma_u$ molecular orbitals. 
Taking into account recent advances in method development for ground and excited states, higher-quality basis sets, and the increase in computation power, it is now possible to deepen our knowledge on the ground and electronic excited states of \ce{Yb2} and provide more reliable benchmark data for potential energy curves (both ground and excited states) that can be exploited in future experimental manipulations of \ce{Yb2}.
The main goal of our work is, thus, to provide a reliable description of the ground and electronic excited states potential energy surfaces of the \ce{Yb2} dimer using modern wave function-based quantum chemistry methods.

\section*{\sffamily \Large METHODOLOGY}

\subsection*{\sffamily \large Basis sets and scalar relativity}
In all our calculations, we used the all-electron atomic natural orbital relativistic correlation consistent (ANO-RCC) basis sets available in the OpenMolcas program package~\cite{}, optimized specifically for the 2-nd order Douglas--Kroll--Hess (DKH) Hamiltonian~\cite{ano-rcc-lanthanides}. 
We employed the triple-$\zeta$ (TZ), quadruple-$\zeta$ (QZ), and ``large'' quality basis sets with the following contraction schemes: $25s22p15d11f4g2h\rightarrow8s7p4d3f2g$, $25s22p15d11f4g2h\rightarrow9s8p5d4f3g$, $25s22p15d11f4g2h\rightarrow11s10p8d7f4g$, respectively. 
The most accurate calculations for the ground state included a fully uncontracted ANO-RCC basis set and the 5-th order DKH Hamiltonian. 
We should note that the quality of the calculated potential energy surfaces is not affected by the order of the DKH transformation.   
Thus, for all other electronic structure methods, scalar relativistic effects were accounted for by the second order Douglas--Kroll--Hess Hamiltonian (DKH2)~\cite{dkh1, dkh2}.

\subsection*{\sffamily \large pCCD-based methods}
All pair Coupled Cluster Doubles (pCCD)~\cite{geminals_7,geminals_6} calculations, also known as the Antisymmetric Product of 1-reference orbital Geminal (AP1roG), were performed using our locally developed~\textsc{PIERNIK}~\cite{piernik100} software package. 
The pCCD ansatz can be written as
\begin{equation}\label{eq:ap1rog}
|{\rm pCCD}\rangle = \exp \left (  \sum_{i=1}^{\rm occ} \sum_{a=1}^{\rm virt} t_i^a a_a^{\dagger}  a_{\bar{a}}^{\dagger}a_{\bar{i}} a_{i}  \right )| 0 \rangle = e^{T_{\rm p}} | 0 \rangle,
\end{equation}
where $a^\dagger_p$ and $a_p$ ($a^\dagger_{\bar{p}}$ and $a_{\bar{p}}$) are the electron creation and annihilation operators for $\alpha$ ($\beta$) electrons and $| 0 \rangle$ is some independent-particle wave function (for instance, the Hartree--Fock (HF) determinant).
In eq.~\eqref{eq:ap1rog}, $\{t_i^a\}$ are the electron-pair amplitudes and $\hat{T}_{\rm p} = \sum_{i=1}^{\rm occ} \sum_{a=1}^{\rm virt} t_i^a a_a^{\dagger}  a_{\bar{a}}^{\dagger}a_{\bar{i}} a_{i} $ is the electron-pair excitation operator that excites an electron pair from an occupied orbital $(i\ib)$ to a virtual orbital $(a\ab)$ with respect to $| 0 \rangle$. 
In all pCCD and post-pCCD calculations, we used different sets of orbitals: canonical Hartree--Fock orbitals and variationally optimized pCCD orbitals (denoted as voo-pCCD)~\cite{OO-AP1roG,PS2-AP1roG,AP1roG-JCTC}. 
Furthermore, voo-pCCD calculations were constraint to the \ce{D_{2h}} and \ce{C_{2v}} point group symmetry, respectively.  
The missing dynamical energy correction on top of pCCD/voo-pCCD was included via a linearized coupled cluster correction~\cite{geminals_lcc_2015}, denoted as pCCD-LCCSD. 

Moreover, three variants of the equation of motion (EOM) coupled cluster model to target excited states within the pCCD formalism were investigated.
First, the EOM formalism was directly applied on top of the pCCD reference. Single excitations were included \textit{a posteriori} in the EOM ansatz, that is, the linear excitation operator of the EOM formalism is limited to pair and single excitations. 
This approach is denoted as EOM-pCCD+S~\cite{eom-pccd,eom-pccd-erratum}. 
The second model includes single excitations also in the coupled cluster reference function. These single excitations are included on top of the pCCD reference function. This method is labeled as pCCD-CCS, while the excited state extension is abbreviated as EOM-pCCD-CCS.
Finally, in the most accurate EOM variant, the pCCD reference was replaced by the pCCD-LCCSD function, resulting in the EOM-pCCD-LCCSD approach~\cite{eom-pccd-lccsd,ola-book-chapter-2019}.

\subsection*{\sffamily \large EOM-CCSD/CCSD(T)}
All CCSD(T)~\cite{DEEGAN1994321} and EOM-CCSD~\cite{molpro-eom-ccsd} calculations were carried out in the \textsc{Molpro2012} software package~\cite{molpro2012,molpro2012_2,casscf-01,casscf-02} using \ce{D_{2h}} point group symmetry. 

\subsection*{\sffamily \large CASSCF/SO-CASPT2}
The Complete Active Space Self Consistent Field (CASSCF)~\cite{Roos_casscf,Siegbahn1981} calculations and Complete Active Space Second-order Perturbation Theory (CASPT2)~\cite{caspt21,caspt22,MS-CASPT2} calculations were performed in the \ce{D_{2h}} point group symmetry using the OpenMolcas (version 17.0) software package~\cite{Molcas6, Molcas7, Molcas-code, Molcas8}. 
The CASSCF wave functions were used to calculate multistate CASPT2 energy corrections, where the ionization potential-electron
affinity (IPEA) shifted H$_0$ Hamiltonian~\cite{caspt2-h0} was applied with an imaginary shift set to 0.25. 
The spin--orbit (SO) interaction effects (in the Atomic Mean Field Approximation\mbox{~\cite{AMFI_1,AMFI_2,AMFI_3}}) were calculated using the Restricted Active Space State Interaction (RASSI) approach,\mbox{~\cite{rassi}} where the energy correction due to dynamical correlation was included in an approximate manner by dressing the diagonal elements of the spin-orbit Hamiltonian by the CASPT2 energies.
Throughout this work, we utilized the state averaged CASSCF approach combined with the subsequent RASSI/SO calculations with the same active space sizes as in CASSCF. 
In the Yb atomic calculations, we employed two active space variants: CAS(2,4)SCF with the 6s and 6p orbitals correlated and CAS(2,9)SCF augmented by the additional 5d orbitals. 
For the \ce{Yb2} molecule, we performed CAS(4,8)SCF calculations which comprised occupied \ce{\sigma_u} (Yb \ce{6s}) and \ce{\sigma_g} (Yb \ce{6s}) orbitals as well as virtual \ce{\sigma_g} (Yb \ce{6p_z}), \ce{\sigma_u} (Yb \ce{6p_z}), \ce{\pi_u} (Yb \ce{6p_x}/\ce{p_y}), and \ce{\pi_g} (Yb \ce{6p_x}/\ce{p_y}) molecular orbitals.

\subsection*{\sffamily \large Fitting procedure}
All potential energy curves were obtained from a polynomial fit of 8-th order.
The corresponding spectroscopic constants (equilibrium bond length (r$_{\rm e}$) and harmonic vibrational frequency ($\omega_{\rm e}$)) were calculated based on those fitted potential energy curves. 
Specifically, the harmonic vibrational frequencies ($\omega_{\rm e}$) were determined numerically using the five-point finite difference stencil~\cite{Abramowitz} and the average mass of ytterbium, that is, 173.045.~\cite{spectro-data-2005}  
The potential energy depth (D$_{\rm e}$) was evaluated as the difference between the atomic limit and the minimum energy of a given potential energy curve.
Note that in pCCD-based calculations employing \ce{D_{2h}} point group symmetry the dissociation energies were estimated by adding the corresponding atomic excitation energies to the dissociation limit of the dimer calculation.
This step had to be performed due to size-consistency problems in the pCCD reference function. If the orbitals are allowed to relax freely, the size-consistency error is eliminated and the dissociation limits of the dimer calculation numerically agree with the atomic ground and excited state energies.
\section*{\sffamily \Large RESULTS and DISCUSSION}

\subsection*{\sffamily \large The electronic structure of the ytterbium atom}
Ytterbium is a closed-shell atom described by the ground-state \ce{^1S_0} term. 
Its valence electronic configuration is characterized by the fully occupied \ce{4f} and \ce{6s} subshells and the low-lying unoccupied \ce{5p}, \ce{5d}, and \ce{7s} subshells. 
Despite its closed-shell electronic nature, the ytterbium atom has a rather complex electronic structure, which manifests itself in close-lying potential energy levels~\cite{spectro-data-2005,nist-yb}. 
Specifically, in the range of 17 000 to 50 000 cm$^{-1}$, there is a large number of quasi-degenerate states that are characterized by electron transfer not only from the occupied \ce{6s} to the unoccupied \ce{5p}, \ce{5d}, \ce{7s}, and \ce{8s} orbitals, but also from the occupied \ce{4f} (7-fold) semi-core orbitals. 
This peculiar electronic structure leads to a large number of low-lying excited states and a very dense electronic spectrum, for some of which atomic term symbols are difficult to assign~\cite{spectro-data-2005,nist-yb}.      

In this work, we focus on the low-lying energy levels of the Yb atom arising from the occupied \ce{6s} to the unoccupied \ce{5p} and \ce{5d} orbitals. 
Specifically, we use the experimentally determined energy levels available in Refs.~\citenum{spectro-data-2005,nist-yb} as a starting point to assess the accuracy of different quantum chemistry methods.  
Since not all quantum chemistry methods used in this work are directly applicable to triplet excited states (EOM-based approaches), we first focus on the lowest-lying \ce{^1D} ($6s\rightarrow5d$) and \ce{^1P} ($6s\rightarrow6p$) energy levels of Yb. 
The corresponding results are summarized in Table~\ref{tbl:yb-singlets}. 
\begin{table}[h!]
\begin{adjustbox}{width=\columnwidth,center}
\begin{tabular}{|c|c|c|c|c|c|c|c|c|}\hline
\multicolumn{1}{|c|}{}&
\multicolumn{1}{c|}{}&
\multicolumn{7}{c|}{\textbf{Spin--free levels $[$cm$^{-1}]$}} \\\cline{3-9}

\multicolumn{1}{|c|}{\textbf{Main config.}}&
\multicolumn{1}{c|}{\textbf{Term}}&
\multicolumn{4}{c|}{\textbf{EOM}}&
\multicolumn{2}{c|}{\textbf{CAS(2,9)}}&
\multicolumn{1}{c|}{\textbf{Exp.$^{*}$ }}\\\cline{3-9}

 && \textbf{pCCD+S} & \textbf{pCCD-CCS} & \textbf{pCCD-LCCSD} & \textbf{CCSD} &\textbf{CASSCF} &\textbf{CASPT2} &\textbf{Ref.}~\citenum{spectro-data-2005}\\ \hline 
 4f$^{14}$6s6p            &$^1$P& 29 131 & 29 127 & 26 452 & 25 826 &24 672 &25 007 & 24 964 \\\hline 
 4f$^{14}$6s5d            &$^1$D& 29 414 & 29 355 & 29 303 & 30 182 &27 840 &27 574 & 27 628 \\\hline 
\end{tabular}
\end{adjustbox}
\begin{tablenotes}
        \item[a] $^*$ 
{\small The empirical positions 24 964 cm$^{-1}$ and 18 903 cm$^{-1}$ respectively of spin-free $^1$P and $^3$P states were determined from experimental positions of $^1$P$_1$ and $^3$P$_{0,1,2}$ states using the spin-orbit Hamiltonian in Ref.~\citenum{mies_1978} and take into account the mutual repulsion between $^3$P$_1$ and $^1$P$_1$ states.
Analogously, the empirical positions of spin-free $^1$D and $^3$D states are 27 628 cm$^{-1}$ and 24 958 cm$^{-1}$, respectively, and account for the repulsion between the $^1$D$_2$ and $^3$D$_2$ states.}
\end{tablenotes}
\caption{Singlet electronic energy levels of the Yb atom calculated from different quantum chemistry methods using the TZ-ANO-RCC basis set (in cm$^{-1}$). The energy of the $^1$S ground state (electronic configuration 4f$^{14}$6s$^2$) equals zero for all theoretical models and experiment.}
\label{tbl:yb-singlets} 
\end{table}

All excited state methods correctly place the $^1$P state below $^1$D; nonetheless, the splitting between these two states differs for all investigated approaches. 
All considered EOM-based theories overestimate the energy of the $^1$D level by approximately 2 000 cm$^{-1}$. 
The $^1$P state is rather accurately predicted by the standard EOM-CCSD method, followed by EOM-pCCD-LCCSD.
For more simplified EOM models, however, larger deviations from the reference value are observed (differences amount to 3 000 cm$^{-1}$). 
This results in underestimated energy splittings between the $^1$P and $^1$D terms. 
On the other hand, spin--free CASSCF and CASPT2 electronic spectra match very well the experimental energy levels mentioned in Table~\ref{tbl:yb-singlets}. 
The overall deviations do not exceed 300 and 100 cm$^{-1}$ for CASSCF and CASPT2, respectively.  

The Yb energy levels obtained from SO-CAS(2,4)PT2 and SO-CAS(2,9)PT2 combined with various basis set sizes are presented in Tables~\ref{tbl:yb-cas24} and~\ref{tbl:yb-cas29}. 
Both active spaces qualitatively reproduce the experimental energy levels of the Yb atom with respect to the energetic order and magnitude of spin-orbit splittings. 
The largest differences between the CAS(2,4) and CAS(2,9) variants are observed for the $^1$P state, which is underestimated by about 2 000 cm$^{-1}$ in the smaller CAS. 
This shift in energy highlights the importance of unoccupied d-type orbitals in post-Hartree--Fock calculations, similar to the \textit{double d-shell effect} in 3d transition metal chemistry.~\cite{Roos2008} 
In both active space calculations, dynamic energy corrections seem to be important and amount to 500--2 000 cm$^{-1}$ for a given energy level. 
Finally, we should note that the quality of excitation energies in the Yb atom seems to be rather insensitive to the basis set quality. 
Only minor changes in the order of states are observed and amount to a few hundreds of cm$^{-1}$ (cf. Table~\ref{tbl:yb-cas29}). 
These observations point to a well-balanced, good quality ANO-RCC basis set for valence atomic properties of the Yb atom.        
\begin{table}[h!]
\begin{adjustbox}{width=\columnwidth,center}
\begin{tabular}{|c|c|c|c|c|c|c|c|c|}\hline
\multicolumn{1}{|c|}{\textbf{\Lower{Basis}}}&
\multicolumn{1}{c|}{\textbf{\Lower{Main config.}}}&
\multicolumn{1}{c|}{\textbf{\Lower{Term}}}&
\multicolumn{3}{c|}{\textbf{Spin--free levels $[$cm$^{-1}]$}}&
\multicolumn{3}{c|}{\textbf{Spin--orbit levels $[$cm$^{-1}]$}}\\ \cline{4-9}
 &&&\textbf{CAS(2,4)SCF}&\textbf{CASPT2}&\textbf{Exp.}$^*$~\citenum{spectro-data-2005}&\textbf{J}&\textbf{SO-CASPT2}&\textbf{Exp.}~\citenum{spectro-data-2005}\\ \hline
 
 \multirow{5}{*}{\rotatebox{90}{\footnotesize{TZ-ANO-RCC}}}
 &4f$^{14}$6s6p            &$^3$P&16 304&17 123&18 903&0&15 965   &17 288\\
 &                       &   & &               &      &1&16 523   &17 992 \\ 
 &                       &   & &               &      &2&17 704   &19 710\\ \cline{2-9} 
 &4f$^14$6s6p            &$^1$P&29 608&28 011  &24 964&1&28 041   &25 068\\\hline \hline 

 \multirow{5}{*}{\rotatebox{90}{\footnotesize{QZ-ANO-RCC}}}
 &4f$^{14}$6s6p            &$^3$P&16 315&17 207&18 903&0&16 074 &17 288\\
 &                       &   & &               &      &1&16 628 &17 992 \\ 
 &                       &   & &               &      &2&17 779 &19 710\\ \cline{2-9} 
 &4f$^14$6s6p            &$^1$P&28 966&27 430& 24 964 &1&27 453 &25 068\\\hline \hline

 \multirow{5}{*}{\rotatebox{90}{\footnotesize{ large-ANO-RCC}}}
 &4f$^{14}$6s6p            &$^3$P&16 315&17 256&18 903&0&16 120&17 288\\
 &                       &   & &               &      &1&16 675&17 992\\ 
 &                       &   & &               &      &2&17 828&19 710\\ \cline{2-9} 
 &4f$^{14}$6s6p            &$^1$P&28 512&27 025&24 964&1&27 046&25 068\\\hline \hline
\end{tabular}
\end{adjustbox}
\begin{tablenotes}
        \item[a]  
{\small $^*$ See the footnote below Table~\ref{tbl:yb-singlets}.}
\end{tablenotes}

\caption{Electronic energy levels of the Yb atom calculated from the SO-CAS(2,4)PT2 approach and different quality ANO-RCC basis sets (in cm$^{-1}$). The energy of the $^1$S ground state (electronic configuration 4f$^{14}$6s$^2$) equals zero for all theoretical models and experiment.}  
\label{tbl:yb-cas24}
\end{table}

\begin{table}[h!]
\scalebox{0.7}{%
\begin{adjustbox}{width=\columnwidth,center}
\centering
\begin{tabular}{|c|c|c|c|c|c|c|c|c|}\hline
\multicolumn{1}{|c|}{\textbf{\Lower{Basis}}}&
\multicolumn{1}{c|}{\textbf{\Lower{Main config.}}}&
\multicolumn{1}{c|}{\textbf{\Lower{Term}}}&
\multicolumn{3}{c|}{\textbf{Spin--free levels $[$cm$^{-1}]$}}&
\multicolumn{3}{c|}{\textbf{Spin--orbit levels $[$cm$^{-1}]$}}\\ \cline{4-9}
 &&&\textbf{CAS(2,4)SCF}&\textbf{CASPT2}&\textbf{Exp.}$^*$~\citenum{spectro-data-2005}&\textbf{J}&\textbf{SO-CASPT2}&\textbf{Exp.}~\citenum{spectro-data-2005}\\ \hline
 
 \multirow{9}{*}{\rotatebox{90}{TZ-ANO-RCC}}
 &4f$^{14}$6s6p            &$^3$P &15 497&16 694&18 903&0&15 528&17 288\\
 &                       &   & &                &      &1&16 064&17 992\\ 
 &                       &   & &                &      &2&17 276&19 710\\ \cline{2-9} 
 &4f$^{14}$6s5d            &$^3$D&26 895&25 684 &24 958&1&25 149&24 489\\ 
 &                       &     &&               &      &2&25 443&24 752\\ 
 &                       &     &&               &      &3&26 040&25 271\\ \cline{2-9}
 &4f$^14$6s6p            &$^1$P&24 672& 25 007  &24 964&1&25 053&25 068\\\cline{2-9}
 &4f$^14$6s5d            &$^1$D&27 840&27 574   &27 628&2&27 690&27 678\\\hline \hline 

 \multirow{9}{*}{\rotatebox{90}{QZ-ANO-RCC}}
 &4f$^{14}$6s6p            &$^3$P &15 416&16 862&18 903&0& 15 720  &17 288\\
 &                       &   & &                &      &1&16 249&17 992\\ 
 &                       &   & &                &      &2&17 433&19 710\\ \cline{2-9} 
 &4f$^{14}$6s5d            &$^3$D&26 573&25 124 &24 958&1&24 614&24 489\\ 
 &                       &     &&               &      &2&24 899&24 752\\ 
 &                       &     &&               &      &3&25 474&25 271\\ \cline{2-9}
 &4f$^14$6s6p            &$^1$P&24 275&24 793   &24 964&1&24 834&25 068\\\cline{2-9}
 &4f$^14$6s5d            &$^1$D&27 485&27 142   &27 628&2&27 265&27 678\\\hline \hline 

 \multirow{9}{*}{\rotatebox{90}{large-ANO-RCC}}
 &4f$^{14}$6s6p            &$^3$P &15 381&16 997&18 903&0&15 870&17 288\\
 &                       &   & &                &      &1&16 399&17 992\\ 
 &                       &   & &                &      &2&17 560&19 710\\ \cline{2-9} 
 &4f$^{14}$6s5d          &$^3$D&26 414&24 423   &24 958&1&23 909&24 489\\ 
 &                       &     &&               &      &2&24 202&24 752\\ 
 &                       &     &&               &      &3&24 747&25 271\\ \cline{2-9}
 &4f$^{14}$6s6p            &$^1$P&24 170&24 848 &24 964&1&24 888&25 068\\\cline{2-9}
 &4f$^{14}$6s5d            &$^1$D&27 191&26 373 &27 628&2&26 500&27 678\\\hline \hline 

\end{tabular}
\end{adjustbox}
}
\begin{tablenotes}
        \item[a] $^*$ 
{\small See the footnote below Table~\ref{tbl:yb-singlets}.}
\end{tablenotes}
\caption{Electronic energy levels of the Yb atom calculated from the SO-CAS(2,9)PT2 approach and different quality ANO-RCC basis sets (in cm$^{-1}$).
The energy of the $^1$S ground state (electronic configuration 4f$^{14}$6s$^2$) equals zero for all theoretical models and experiment.}
\label{tbl:yb-cas29}
\end{table}

\subsection*{\sffamily \large Towards a reliable and accurate ground state potential energy surface for the \ce{Yb2} dimer}
When the two Yb atoms approach each other, they create a weakly-bonded, van der Waals-type complex.~\cite{andrews-1sigmau-yb2,dolg-yb2,dolg-yb-clusters,Buchachenko2007} 
It is well-known that such weakly-bonded compounds tend to be extremely sensitive to the quality of the atomic basis set and the (approximate) dynamic energy correction.~\cite{a24} 
Thus, we first scrutinize the ground state potential energy surface obtained from the CCSD(T) approach before benchmarking various electron correlation methods for both ground and excited states. 
\subsubsection*{\sffamily \normalsize Reference ground-state potential energy curve}
Table~\ref{yb2-gs-spect-const} summarizes the influence of the number of correlated electrons on the quality of the CCSD and CCSD(T) potential energy surfaces, respectively, including their spectroscopic constants (optimal bond lengths ($\mathrm {r_e}$), harmonic vibrational frequencies ($\omega_e$), and potential energy depths ($\mathrm {D_e}$)). 
To minimize the basis set superposition error~\cite{counter-poise-correction} we employed the all-electron uncontracted ANO-RCC basis set.   
Our calculations suggest that it is necessary to correlate all occupied orbitals starting from the fourth atomic shell of the Yb atom (see Table~\ref{yb2-gs-spect-const}), that is, 84 electrons of the Yb dimer have to be correlated in a calculation. 
Correlating additional core electrons does not considerably change the spectroscopic constants of \ce{Yb2}.  
Our most accurate CCSD(T) prediction yields an optimal bond length of $\mathrm {r_e}=8.814$ bohr, a harmonic vibrational frequencies of $\omega_e=21$ cm$^{-1}$, and a potential energy depth of $\mathrm {D_e}=579$ cm$^{-1}$. 
The computed harmonic vibrational frequency matches the experimental value of 22 cm$^{-1}$ measured by Goodfriend~\cite{goodfriend-yb2}. 
It is important to note that restricting the number of correlated electrons to 32 or less shifts the optimal bond length towards longer inter-atomic distances and overestimates the potential energy depth. 
Neglecting contributions from triply-excited determinants (as in CCSD) elongates the optimal bond length by approximately 0.4 bohr and lowers the vibrational frequencies and potential energy depth. 
The latter is most significantly affected by the lack of triple excitations in the cluster operator, where the differences in $D_e$ between CCSD and CCSD(T) amount to 250--300 cm$^{-1}$ (a difference of approximately 40\%, see also Table~\ref{yb2-gs-spect-const}). 
Analysis of the t$_1$ diagnostic~\cite{lee_1989} in CCSD along potential energy surface shows a single-reference nature of the Yb$_2$ ground-state (values in the order of 0.02). 
Finally, we should mention that our new best estimate for the potential energy depth, $\mathrm {D_e}=579$ cm$^{-1}$, is lower than the recently reported value of $\mathrm {D_e}=786$ cm$^{-1}$ by Mosyagin and coworkers~\cite{titov-yb2}, who employed a smaller basis set and different approaches to electron correlation and relativistic effects. 
\begin{table}[h!]
\centering
\scalebox{0.7}{%
\begin{adjustbox}{width=\columnwidth,center}
\begin{tabular}{|c|c|c|c|c|c|}\hline 
 method& correlated occupied orbitals &$\mathrm {N_e}$ & $\mathrm {r_e [a_0]}$& $\mathrm {\omega_e [cm^{-1}]}$ & $\mathrm {D_e [cm^{-1}]}$\\ \hline 
 CCSD    &6s                                                 &4  &9.438&17&365\\ 
 CCSD    &5s, 5p, 6s                                         &20 &9.373&16&342\\ 
 CCSD    &4f, 6s                                             &32 &9.293&16&345\\
 CCSD    &5p, 4f, 6s                                         &44 &9.281&16&328\\
 CCSD    &5s, 5p, 4f, 6s                                     &48 &9.277&16&336\\
 CCSD    &4s, 4p, 4d, 5s, 5p, 4f, 6s                         &84 &9.260&16&336\\ 
 CCSD    &3s, 3p, 3d, 4s, 4p, 4d, 5s, 5p, 4f, 6s             &120&9.266&16&336\\ 
 CCSD    &1s, 2s, 2p, 3s, 3p, 3d, 4s, 4p, 4d, 5s, 5p, 4f, 6s &140&9.265&16&336\\ \hline 
 CCSD(T) &6s                                                 &4  &9.050&22&645\\ 
 CCSD(T) &5s, 5p, 6s                                         &20 &8.968&21&592\\ 
 CCSD(T) &4f, 6s                                             &32 &8.874&21&591\\
 CCSD(T) &5p, 4f, 6s                                         &44 &8.822&21&579\\
 CCSD(T) &5s, 5p, 4f, 6s                                     &48 &8.810&21&585\\
 CCSD(T) &4s, 4p, 4d, 5s, 5p, 4f, 6s                         &84 &8.815&21&580\\ 
 CCSD(T) &3s, 3p, 3d, 4s, 4p, 4d, 5s, 5p, 4f, 6s             &120&8.814&21&579\\ 
 CCSD(T) &1s, 2s, 2p, 3s, 3p, 3d, 4s, 4p, 4d, 5s, 5p, 4f, 6s &140&8.814&21&579\\ \hline
\end{tabular}
\end{adjustbox}
}
\caption{\label{yb2-gs-spect-const} CCSD and CCSD(T) spectroscopic constants for the~\ce{Yb2} X $\mathrm{^1\Sigma_g^+}$ state using uncontracted ANO-RCC basis set and a varying number of active (occupied) orbitals and thus correlated electrons. 
N$_{\text{corr}}$ denotes the number of correlated electrons, $\mathrm {r_e}$ the equilibrium bond length, $\mathrm {\omega_e}$ the vibrational frequency, and  $\mathrm {D_e}$ the potential depth, respectively. The occupied orbitals not included in the set of correlated orbitals (second column) were kept frozen during CCSD(T) calculations. All virtual orbitals were correlated. }
\end{table}
\subsubsection*{\sffamily \normalsize Assessing the accuracy of conventional and unconventional quantum chemistry approaches in modeling the ground state potential energy surface}
The CCSD potential energy curve can be used as a reference to evaluate the reliability of simplified coupled cluster methods that include at most double excitations. 
Table~\ref{tbl:yb2-gs-methods} lists the spectroscopic constants obtained from various quantum chemistry methods and a given basis set. 
Including an LCCSD dynamic energy correction on top of the pCCD reference wave-function is indispensable for obtaining qualitatively correct bond lengths.
Furthermore, the potential energy depth heavily depends on the type of orbitals used in the pCCD reference calculations (canonical Hartree--Fock or variationally optimized orbitals imposing C$_{2v}$ and D$_{2h}$ point group symmetry). 
The best agreement of all pCCD-based methods with CCSD (as well as with CCSD(T) reference) data is obtained when the point group symmetry is lowered and the orbitals are thus allowed to (partially) localize in the dimer calculation.
Note that orbital optimization in pCCD typically involves localization.~\cite{geminals_9,geminals_3}
CAS(4,8)PT2 results in overly bond lengths and underestimated low vibrational frequencies compared to CCSD calculations (employing the same basis set). 
Increasing the basis set size in CAS(4,8)PT2 improves the agreement between CCSD and CASPT2 data. 
The slower convergence of the second-order perturbation theory approach with basis set size for weakly interacting systems is not surprising and has been reported earlier in the literature~\cite{Helgaker-basis-convergence-water-1997}.
As to be expected, inclusion of spin--orbit coupling does not significantly affect the quality of the CAS(4,8)PT2 ground-state potential energy curve. 
To this end, we can conclude that both the pCCD-LCCSD($C_{2v}$) and CAS(4,8)PT2 methods are promising alternatives for modeling excited state potential energy curves in the Yb$_2$ dimer.  
\begin{table}[h!]
\centering
\scalebox{0.7}{%
\begin{adjustbox}{width=\columnwidth,center}
\begin{tabular}{|c|c|c|c|c|c|}\hline 
 method& $\mathrm {N_e}$ &basis& $\mathrm {r_e [a_0]}$& $\mathrm {\omega_e [cm^{-1}]}$ & $\mathrm {D_e [cm^{-1}]}$\\ \hline 
CCSD                  & 48    &TZ-ANO-RCC      &8.684&24&956\\
pCCD                  & 48    &TZ-ANO-RCC      &10.004&14&361\\
pCCD($D_{2h}$)        & 48    &TZ-ANO-RCC      &8.604&27&1 046\\
pCCD($C_{2v}$)        & 48    &TZ-ANO-RCC      &13.674&22&22\\
pCCD-LCCSD            & 48    &TZ-ANO-RCC      &8.102&36&1 875\\
pCCD-LCCSD($D_{2h}$)  & 48    &TZ-ANO-RCC      &7.852&42&2 609\\
pCCD-LCCSD($C_{2v}$)  & 48    &TZ-ANO-RCC      &8.467&26&785\\

CAS(4,8)PT2           & 4/32  &TZ-ANO-RCC      &11.381 & 13 & 280 \\ 
CAS(4,8)PT2           & 4/32  &QZ-ANO-RCC      &10.998 & 11 & 256\\ 
CAS(4,8)PT2           & 4/32  &large-ANO-RCC   & 8.837 & 19 & 369\\ \hline

SO-CAS(4,8)PT2        & 4/32  &TZ-ANO-RCC      &10.751 & 11 & 221 \\ 
SO-CAS(4,8)PT2        & 4/32  &QZ-ANO-RCC      &10.463 &  9 & 259\\ 
SO-CAS(4,8)PT2        & 4/32  &large-ANO-RCC   & 8.519 & 24 & 409\\ \hline

\end{tabular}
\end{adjustbox}
}
\caption{\label{tbl:yb2-gs-methods} Spectroscopic constants for the~\ce{Yb2} X $\mathrm{^1\Sigma_g^+}$ state from different quantum chemistry methods. 
N$_{\text{corr}}$ denotes the number of correlated electrons, $\mathrm {r_e}$ equilibrium bond length, $\mathrm {\omega_e}$ vibrational frequency, and  $\mathrm {D_e}$ potential depth, respectively. }
\end{table}
 
\subsection*{\sffamily \large \ce{Yb2} excited-state properties} 
Examination of the electronic structure of the Yb atom points to the importance of the 4$f$, 6$s$, 6$p$, and 5$d$ atomic orbitals in the electronic spectrum of Yb$_2$. 
Including all these orbitals in active space calculations is prohibitive and some compromise has to be made. 
A reasonable choice would be to correlate only the 6$s$, 6$p$, and 5$d$ atomic orbitals in molecular Yb$_2$ calculations, that is, performing CAS(4,18)SCF calculations. 
Unfortunately, such an active space is not stable along the potential energy surface, where smooth potential energy curves for all excited states of interest cannot be optimized due to technical difficulties. 
As a consequence, we had to reduce the number of active orbitals and neglected the contributions from 5$d$ orbitals by moving them outside the CAS space (into the external space). 
This results in our CAS(4,8)SCF model that is further used as reference for all Yb$_2$ excited states potential energy curves. 
We should stress that the same active space was used in previous ECP/MRCI calculations~\cite{dolg-yb2}. 
\subsubsection*{\sffamily \normalsize Reference spin-free electronic spectrum}
Table~\ref{tbl:yb2-excit-sf} collects all spin--free spectroscopic constants obtained from CAS(4,8)PT2 using different sizes for the atomic basis set. 
Similar to the ground state calculations, the CAS(4,8)PT2 spectroscopic constants converge very slowly with basis set size. 
In general, incrementing the basis set size shortens bond lengths, marginally increases vibrational frequencies, deepens potential energy depths, and lowers the adiabatic excitation energies. 
The CAS(4,8)PT2 results obtained using the TZ-ANO-RCC basis set qualitatively match the ECP/MRCI~\cite{dolg-yb2} spectroscopic constants (cf. Table~\ref{tbl:yb2-excit-sf}).
The largest discrepancies are observed for the higher-lying $^1\Pi_u$ and $^1\Sigma_g^+$ excited states, where the differences in excitation energies amount to few thousand wave numbers. 
The agreement between ECP/MRCI and all-electron CAS(4,8)PT2 spectroscopic parameters decreases for larger basis sets, indicating the need for decent basis set sizes to reliably describe excited-state potential energy surfaces of Yb$_2$.  
The discrepancies between ECP/MRCI and CAS(4,8)PT2 can be attributed to the different treatment of scalar relativistic and electron correlation effects in both approaches. 
One should keep in mind that any truncated CI approach, such as MRCI, is not rigorously size-extensive and size-consistent, while CASPT2 is, in general, size-extensive and approximately size-consistent. 
Thus, we believe that our CAS(4,8)PT2 results can be considered as new reference data for the excited potential energy curves of the Yb$_2$ dimer. 
We further hope that future experiments on laser induced fluorescence will help to resolve this ambiguity.  
\begin{table}[ht!]
\centering
\scalebox{0.7}{%
\begin{tabular}{|c|c|c|c|c|c|c|}\hline 
State &basis & $\mathrm {r_e [a_0]}$& $\mathrm {\omega_e [cm^{-1}]}$ & $\mathrm {D_e [cm^{-1}]}$ &$\mathrm {T_e [cm^{-1}]}$ & Dissociation limit\\ \hline 
$^3\Pi_g$&TZ-ANO-RCC        &6.955&69&6 255&11 147&$\mathrm{^1S+^3P}$\\
         &QZ-ANO-RCC        &6.890&69&6 640&10 823&$\mathrm{^1S+^3P}$\\
         &large-ANO-RCC     &6.794&73&7 566&10 059&$\mathrm{^1S+^3P}$\\
         &ECP(MRCI)~\cite{dolg-yb2}         &6.680&77&8 065&12 421&$\mathrm{^1S+^3P}$\\\hline

$^1\Pi_g$&TZ-ANO-RCC        &6.751&79&15 776&12 514&$\mathrm{^1S+^1P}$\\
         &QZ-ANO-RCC        &6.688&77&15 684&12 002&$\mathrm{^1S+^1P}$\\
         &large-ANO-RCC     &6.612&80&16 364&11 030&$\mathrm{^1S+^1P}$\\
         &ECP(MRCI)~\cite{dolg-yb2}          &6.546&84&14 841&13 389&$\mathrm{^1S+^1P}$\\\hline
                                                   
$^3\Sigma_u^+$&TZ-ANO-RCC   &7.697&56&4 141&13 261&$\mathrm{^1S+^3P}$\\
         &QZ-ANO-RCC        &7.636&57&4 493&12 970&$\mathrm{^1S+^3P}$\\
         &large-ANO-RCC     &7.578&61&5 315&12 310&$\mathrm{^1S+^3P}$\\
         &ECP(MRCI)~\cite{dolg-yb2}         &7.559&58&5 162&15 325&$\mathrm{^1S+^3P}$\\\hline

$^3\Pi_u$&TZ-ANO-RCC        &8.835&21&425  &16 978&$\mathrm{^1S+^3P}$\\
         &QZ-ANO-RCC        &8.693&22&522  &16 942&$\mathrm{^1S+^3P}$\\
         &large-ANO-RCC     &8.374&31&802  &16 824&$\mathrm{^1S+^3P}$\\
         &ECP(MRCI)~\cite{dolg-yb2}          &8.343&24&1 048&19 438&$\mathrm{^1S+^3P}$\\\hline

$^1\Sigma_u^+$&TZ-ANO-RCC   &7.513&48&8 774 &19 516&$\mathrm{^1S+^1P}$\\
         &QZ-ANO-RCC        &7.307&52&8 906 &18 781&$\mathrm{^1S+^1P}$\\
         &large-ANO-RCC     &7.078&60&10 405&16 989&$\mathrm{^1S+^1P}$\\
         &ECP(MRCI)~\cite{dolg-yb2}          &7.359&53&7 824 &20 406&$\mathrm{^1S+^1P}$\\\hline
                                                   
$^1\Pi_u$&TZ-ANO-RCC        &7.258&71&5 595&22 695&$\mathrm{^1S+^1P}$\\
         &QZ-ANO-RCC        &7.273&67&5 455&22 232&$\mathrm{^1S+^1P}$\\
         &large-ANO-RCC     &7.223&69&6 124&21 270&$\mathrm{^1S+^1P}$\\
         &ECP(MRCI)~\cite{dolg-yb2}          &7.319&56&1 936&26 294&$\mathrm{^1S+^1P}$\\\hline
                                                   
$^1\Sigma_g^+$&TZ-ANO-RCC   &7.664&68&5 111&23 179&$\mathrm{^1S+^1P}$\\
         &QZ-ANO-RCC        &7.461&71&5 008&22 679&$\mathrm{^1S+^1P}$\\
         &large-ANO-RCC     &7.856&49&4 727&22 667&$\mathrm{^1S+^1P}$\\
         &ECP(MRCI)~\cite{dolg-yb2}          &7.529&58&1 613&26 616&$\mathrm{^1S+^1P}$\\\hline
                                                   
\hline
\end{tabular}
}
\caption{\label{tbl:yb2-excit-sf} Adiabatic spin--free electronic spectrum of \ce{Yb2} from CAS(4,8)PT2 using different ANO-RCC basis sets. $\mathrm {r_e}$ denotes the equilibrium bond length, $\mathrm {\omega_e}$ vibrational frequency, $\mathrm {D_e}$ potential depth, and $\mathrm {T_e}$ adiabatic excitation energy, respectively. The $^3\Sigma_g^{+}$ state does not have a minimum and thus its spectroscopic constants are not calculated. 
}
\end{table}

\begin{figure}[h!]
\caption{\label{fig:double-excit-charact}
Contributions from single (red) and double (turquoise) excitations in the first $^1\Sigma_g^{+}$ excited state of Yb$_2$ from different quantum chemistry methods: (a) EOM-CCSD and CAS(4,8)SCF, (b) EOM-pCCD-CCS (with and without orbital optimization), and (c) EOM-pCCD-LCCSD (with and without orbital optimization and different point group symmetries). Results for EOM-pCCD+S are similar to EOM-pCCD-CCS and are thus not shown in the Figure. For CAS(4,8)SCF, the individual contributions are determined from the weights of all active space configurations with coefficients larger than 0.05 only.  
}
\vspace*{-0.5cm}
\begin{center}
\includegraphics[width=0.99\columnwidth]{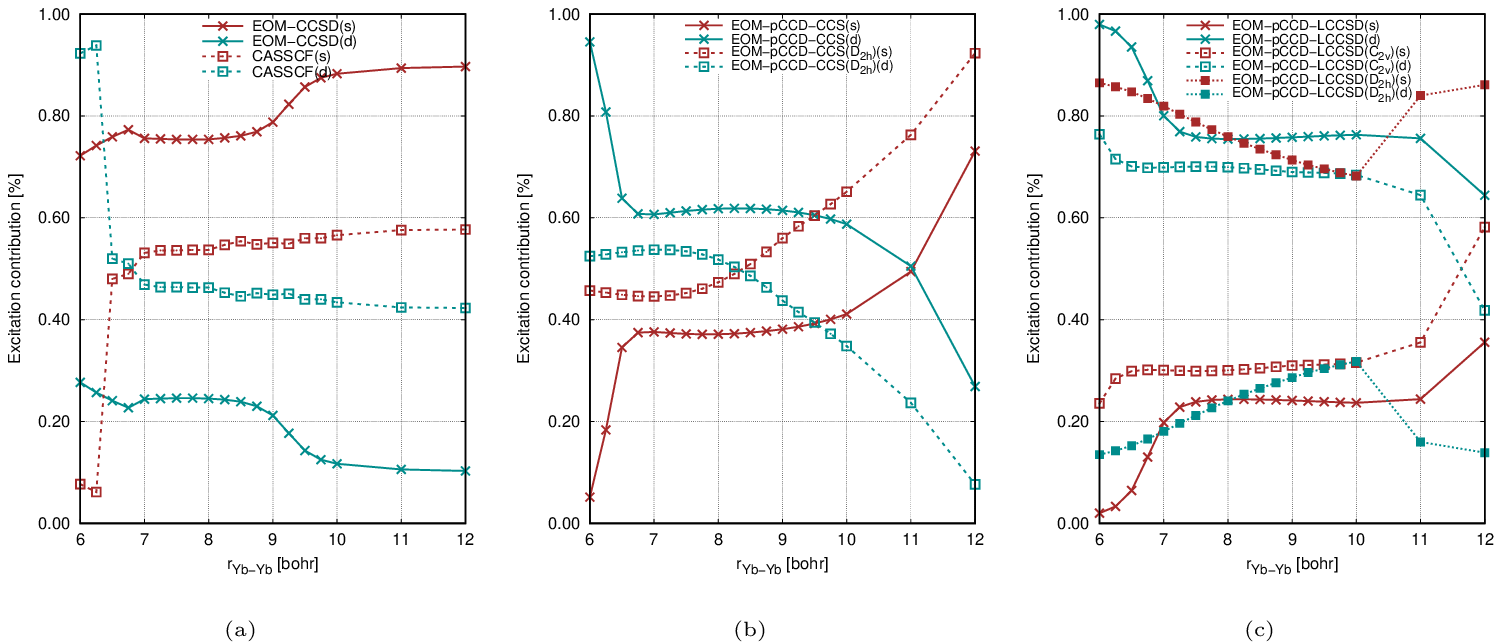}
\end{center}
\vspace*{-0.5cm}
\end{figure}

{\sffamily \small Singlet excitation energies from EOM-based methods}\\
Having generated reference excited-state potential energy curves, we can now assess the accuracy of (simplified) EOM-based methods for singlet excitation energies. 
Table~\ref{tbl:yb2-eom} lists all the spectroscopic constants for excited states obtained from various flavours of EOM methods and their difference with respect to the CAS(4,8)PT2 data. 
In general, the EOM-based excited state energies are overestimated and the potential energy depths underestimated compared to CAS(4,8)PT2 results.  
These differences are smaller when an LCCSD correction is applied on top of pCCD, pushing the EOM-pCCD-LCCSD results very close to EOM-CCSD data. 
Opposite to what we observed for the ground-state, the best performance of EOM-pCCD-LCCSD is achieved when $D_{2h}$ point group symmetry is imposed.
The largest discrepancies between EOM-pCCD-LCCSD and EOM-CCSD can be found for the $^1\Sigma_g^{+}$ excited state. 
Specifically, this excited state features a strong multi-reference nature with a double electron excitation character. 
It is well known that such bi-excited states cannot be correctly described by the standard (single-reference) EOM-CCSD framework.
Recent work on all-trans polyene chains highlights the superiority of EOM-pCCD-based methods to correctly describe double electron excitation energies~\cite{eom-pccd,eom-pccd-erratum,eom-pccd-lccsd}. 
The same is true for the $^1\Sigma_g^{+}$ excited state in the Yb$_2$ dimer. 
Thus, pCCD-based excited state methods outperform the conventional EOM-CCSD formalism. 
Figure~\ref{fig:double-excit-charact} shows the evolution of the bi-excitation character in the $^1\Sigma_g^{+}$ excited state along the potential energy surface.  
For short inter-atomic distances the doubly-excited state has a dominant contribution in all investigated methods, except for EOM-CCSD. 
Moreover, all EOM-pCCD-type approaches feature a similar evolution of the contribution of doubly-excited states along the $^1\Sigma_g^{+}$ Yb$_2$ potential energy surface, which qualitatively agrees with CAS(4,8)SCF results.
We should note that the excitation contributions to the $^1\Sigma_g^{+}$ excited state in EOM-pCCD-LCCSD($C_{2v}$) are reversed in contrast to all remaining pCCD-based methods.
The orbital optimization and (partial) orbital localization thus lowers the bi-excited character in the $^1\Sigma_g^{+}$ state.
While this observed symmetry-breaking worsens equilibrium bond lengths and vibrational frequencies, excitation energies deviate less compared to the CAS(4,8)PT2 reference values.
Finally, we should stress that EOM-pCCD-LCCSD($D_{2h}$) in general outperforms EOM-CCSD in predicting spectroscopic constants for the lowest-lying excited states (difference amount to 2 500 cm$^{-1}$).
\begin{table}[ht!]
\centering
\scalebox{0.7}{%
\begin{adjustbox}{width=\columnwidth,center}
\begin{tabular}{|c|c|c|c|c|c|}\hline 
State &Method  & $\mathrm {r_e [a_0]}$& $\mathrm {\omega_e [cm^{-1}]}$ & $\mathrm {D_e [cm^{-1}]}$ &$\mathrm {T_e [cm^{-1}]}$\\ \hline 
$^1\Pi_g$&EOM-CCSD          &6.643 ($-$0.031)&81 (1)    &13 773 ($-$2 591)&13 035 (2 005)\\
&EOM-pCCD+S                 &7.100    (0.488)&71 ($-$9) &13 596 ($-$2 768)&15 904 (4 874)\\
&EOM-pCCD+S($D_{2h}$)       &6.993    (0.381)&72 ($-$8) &14 347 ($-$2 017)&15 829 (4 799)\\
&EOM-pCCD-CCS               &7.091    (0.479)&71 ($-$9) &13 712 ($-$2 652)&15 775 (4 745)\\
&EOM-pCCD-CCS($D_{2h}$)     &6.992    (0.380)&72 ($-$8) &14 350 ($-$2 014)&15 837 (4 807)\\
&EOM-pCCD-LCCSD             &6.562 ($-$0.050)&82    (2) &15 047 ($-$1 317)&13 276 (2 246)\\
&EOM-pCCD-LCCSD($D_{2h}$)   &6.561 ($-$0.051)&84    (4) &16 046   ($-$318)&13 134 (2 104)\\ 
&EOM-pCCD-LCCSD($C_{2v}$)   &6.588 ($-$0.024)&82    (2) &14 696 ($-$1 668)&13 862 (2 832)\\\hline \hline
                                                   
$^1\Sigma_u^+$&EOM-CCSD     &7.693    (0.615)&47 ($-$13)& 6 788	($-$3 617)&19 996 (3 007)\\
&EOM-pCCD+S                 &8.889 (1.811)&31 ($-$29)   & 7 163 ($-$3 242)&22 337 (5 348)\\
&EOM-pCCD+S($D_{2h}$)       &8.823 (1.745)&33 ($-$27)   & 6 831 ($-$3 574)&23 345 (6 356)\\
&EOM-pCCD-CCS               &8.931 (1.853)&31 ($-$29)   & 7 095 ($-$3 319)&22 392 (5 403)\\
&EOM-pCCD-CCS($D_{2h}$)     &8.821 (1.743)&33 ($-$27)   & 6 851 ($-$3 554)&23 337 (6 348)\\
&EOM-pCCD-LCCSD             &7.575 (0.497)&48 ($-$12)   & 7 682 ($-$2 723)&20 641 (3 652)\\
&EOM-pCCD-LCCSD($D_{2h}$)   &7.448 (0.370)&51 ($-$9)    & 8 647 ($-$1 758)&20 532 (3 543)\\
&EOM-pCCD-LCCSD($C_{2v}$)   &7.441 (0.363)&49 ($-$11)   & 8 075 ($-$2 330)&20 484 (3 495)\\\hline\hline
                                                   
$^1\Pi_u$&EOM-CCSD          &7.488 (0.265)&48 ($-$21)   & 3 610 ($-$2 514)&27 529 (6 259)\\
&EOM-pCCD+S(HF)             &8.128 (0.905)&43 ($-$26)   & 1 689 ($-$4 435)&27 811 (6 541)\\
&EOM-pCCD+S($D_{2h}$)       &7.802 (0.579)&53 ($-$16)   & 2 604 ($-$3 520)&27 572 (6 302)\\
&EOM-pCCD-CCS               &7.995 (0.772)&47 ($-$22)   & 1 785 ($-$4 339)&27 702 (6 432)\\
&EOM-pCCD-CCS($D_{2h}$)     &7.800 (0.577)&54 ($-$15)   & 2 611 ($-$3 513)&27 577 (6 307)\\
&EOM-pCCD-LCCSD             &7.300 (0.077)&63  ($-$6)   & 2 882 ($-$3 242)&25 310 (4 040)\\
&EOM-pCCD-LCCSD($D_{2h}$)   &7.226 (0.003)&67 ($-$2)    & 3 574 ($-$2 550)&25 025 (3 755)\\
&EOM-pCCD-LCCSD($C_{2v}$)   &7.296 (0.073)&62 ($-$7)    & 2 039 ($-$4 085)&25 875 (4 605)\\\hline\hline
    
$^1\Sigma_g^{+}$&EOM-CCSD   &7.757 ($-$0.099)&54 (5) & 1 232 ($-$3 495)&26 683 (4 016)\\ 
&EOM-pCCD+S                 &8.083 (0.227)&59 (10)   & 6 961 (2 234)&22 539 ($-$128)\\
&EOM-pCCD+S($D_{2h}$)       &7.621 ($-$235)&79 (30)  & 7 229 (2 502)&22 947 (280)\\
&EOM-pCCD-CCS               &8.095 (0.239) &67 (18)  & 7 056 (2 329)&22 432 ($-$235)\\
&EOM-pCCD-CCS($D_{2h}$)     &7.620 ($-$0.236)&79 (30)& 7 236 (2 509)&22 951 (284)\\
&EOM-pCCD-LCCSD             &7.749 ($-$0.107)&57  (8)& 7 607 (2 880)&20 716 ($-$1 951)\\
&EOM-pCCD-LCCSD($D_{2h}$)   &7.507 ($-$0.349)&70 (21)& 8 117 (3 390)&21 063 ($-$1 604)\\
&EOM-pCCD-LCCSD($C_{2v}$)   &7.127 ($-$0.729)&72 (23)& 3 661 ($-$1 066)&24 898 ( 2 232)\\\hline\hline
\end{tabular}
\end{adjustbox}
}
\caption{\label{tbl:yb2-eom} Spectroscopic constants for the low-lying adiabatic singlet-excited states of \ce{Yb2} obtained from different EOM-based methods and for the TZ-ANO-RCC basis set. $\mathrm {T_e}$ denotes the adiabatic excitation energy, $\mathrm {r_e}$ the equilibrium bond length, $\mathrm {\omega_e}$ the vibrational frequency, and $\mathrm {D_e}$ the potential depth, respectively. The labels ``($D_{2h}$)'' and ``($C_{2v}$)'' indicate the orbital optimized reference ground state within pCCD, imposing $D_{2h}$ and $C_{2v}$ point group symmetries, respectively. The differences with respect to the reference CAS(4,8)PT2 values (large-ANO-RCC) are given in parenthesis. 
}
\end{table}

\subsubsection*{\sffamily \normalsize Reference spin--orbit electronic spectrum}
\begin{figure}[h!]
\caption{\label{fig:so-caspt2-pes}
SO-CAS(4,8)PT2 electronic spectrum of Yb$_2$ using the large-ANO-RCC basis set. The whole spectrum is divided into contributions from $\Sigma$, $\Pi_u$, and $\Pi_g$ states (from left to right).  
}
\vspace*{-0.5cm}
\begin{center}
\includegraphics[width=0.99\columnwidth]{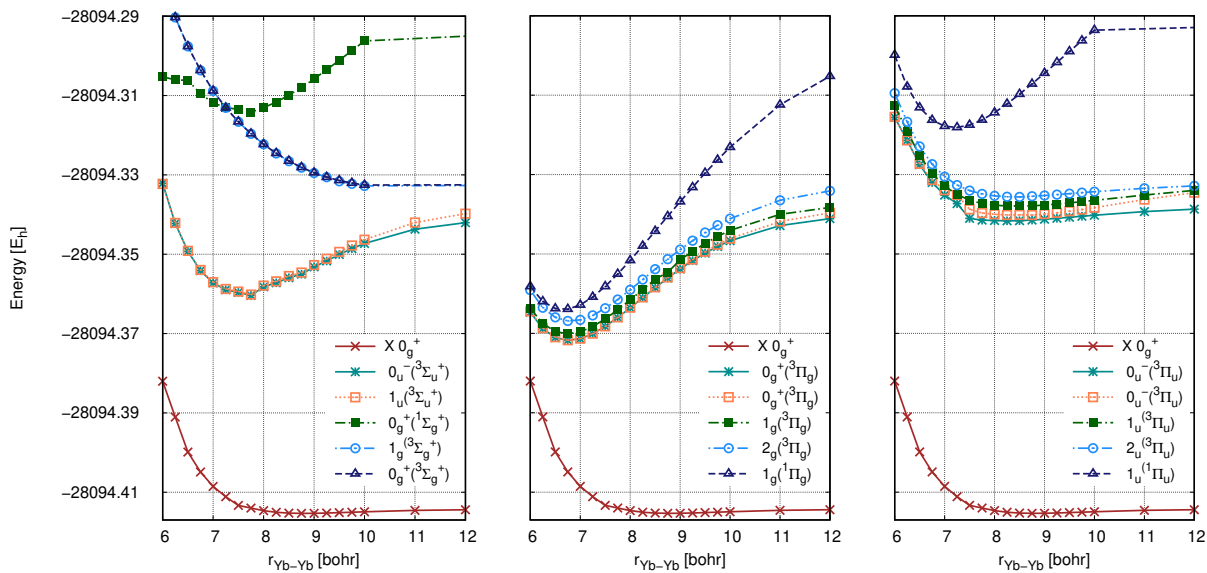}
\end{center}
\vspace*{-0.5cm}
\end{figure}

The reference spin-orbit Yb$_2$ excited-state potential energy surfaces are presented in Figure~\ref{fig:so-caspt2-pes}. 
If spin-orbit coupling is accounted for, the minima of each excited potential energy curve are shifted towards shorter inter-atomic distances.
Furthermore, the whole spectrum is rather dense, especially all states approaching the atomic limits $^1$S+$^3$P$_0$, $^1$S+$^3$P$_1$, and $^1$S+$^3$P$_2$ lie very close to each other. 
Higher lying are the $1_u(^1\Pi_u)$ and $0_g^+(^1\Sigma_g^+)$ states that dissociate into the $^1$S+$^1$P$_1$ atomic limit. 
A distinct feature of the Yb$_2$ electronic spectrum is the presence of 0$_g^+$ excited states originating from the repulsive (spin--free) $^3\Sigma_g^+$ state with a specific shape of the potential energy surface compared to the remaining excited states.
  
The SO-CAS(4,8)PT2 spectroscopic parameters for the excited states in Yb$_2$ are collected in Tables~\ref{tbl:so-caspt2-3p0}--\ref{tbl:so-caspt2-1p1}.  
For convenience, the spectroscopic characteristics of all investigated excited states are grouped into four blocks according to their atomic dissociation limit, Table~\ref{tbl:so-caspt2-3p0} for the $^1$S+$^3$P$_0$, Table~\ref{tbl:so-caspt2-3p1} for the $^1$S+$^3$P$_1$, Table~\ref{tbl:so-caspt2-3p2} for the $^1$S+$^3$P$_2$, and Table~\ref{tbl:so-caspt2-1p1} for the $^1$S+$^1$P$_1$ atomic limits, respectively. 
In the lowest-lying part of the spectrum, our TZ-ANO-RCC results qualitatively match the ECP/SO-MRCI spectroscopic parameters determined by Wang and Dolg~\cite{dolg-yb2}. 
The differences increase, however, when larger basis sets are used within the SO-CAS(4,8)PT2 approach. 
This observation suggests the need for large basis set sizes when targeting excited states in the Yb$_2$ dimer.
In general, the size of the atomic basis set affects all spectroscopic parameters.
A large basis set considerably shortens the optimal bond lengths, lowers the adiabatic excitation energies, and increases potential energy depths.
Harmonic vibrational frequencies are only slightly altered by the choice of the basis set. 

\begin{table}[h!]
\centering
\begin{tabular}{|c|c|c|c|c|c|}\hline 
State &basis & $\mathrm {r_e [a_0]}$& $\mathrm {\omega_e [cm^{-1}]}$ & $\mathrm {D_e [cm^{-1}]}$ &$\mathrm {T_e [cm^{-1}]}$ \\ \hline  
0$_g^{-}$($^3\Pi_{g}$$)$      &TZ-ANO-RCC        &6.938&69&5 627&10 589\\
                              &QZ-ANO-RCC        &6.874&70&6 038&10 294\\
                              &large-ANO-RCC     &6.778&73&6 968& 9 561\\
                              &ECP(MRCI)~\cite{dolg-yb2}         &6.969&65&4 759&13 147\\\hline \hline
                                                   
0$_u^{-}$($^3\Sigma_{u}^+$$)$ &TZ-ANO-RCC        &7.693&54&3 132&13 084\\
                              &QZ-ANO-RCC        &7.640&56&3 506&12 826\\
                              &large-ANO-RCC     &7.568&60&4 332&12 197\\
                              &ECP(MRCI)~\cite{dolg-yb2}        &7.544&54&2 662&15 244\\\hline \hline

\hline
\end{tabular}
\caption{\label{tbl:so-caspt2-3p0} SO-CAS(4,8)PT2 adiabatic electronic states of \ce{Yb2} dissociating into the $^1$S+$^3$P$_0$ atomic limit. $\mathrm {r_e}$ denotes the equilibrium bond length, $\mathrm {\omega_e}$ the vibrational frequency, $\mathrm {D_e}$ the potential depth, and $\mathrm {T_e}$ the adiabatic excitation energy, respectively. 
}
\end{table}

\begin{table}[h!]
\centering
\begin{tabular}{|c|c|c|c|c|c|}\hline 
State &basis & $\mathrm {r_e [a_0]}$& $\mathrm {\omega_e [cm^{-1}]}$ & $\mathrm {D_e [cm^{-1}]}$ &$\mathrm {T_e [cm^{-1}]}$ \\ \hline  

0$_g^{+}(^3\Pi_{g})$          &TZ-ANO-RCC        &6.939&69&6 173&10 600\\
                              &QZ-ANO-RCC        &6.877&70&6 584&10 303\\
                              &large-ANO-RCC     &6.784&73&7 508& 9 578\\
                              &ECP(MRCI)~\cite{dolg-yb2}         &6.971&66&5 404&13 147\\\hline \hline
                                                   
1$_g(^3\Pi_{g})$              &TZ-ANO-RCC        &6.920&69&5 775&10 998\\
                              &QZ-ANO-RCC        &6.877&70&6 584&10 303\\
                              &large-ANO-RCC     &6.753&72&7 154& 9 933\\
                              &ECP(MRCI)~\cite{dolg-yb2}         &6.984&65&4 839&13 711\\\hline \hline
                                                  
1$_u(^3\Sigma_{u}^+)$         &TZ-ANO-RCC        &7.681&55&3 618&13 155\\
                              &QZ-ANO-RCC        &7.633&57&3 998&12 889\\
                              &large-ANO-RCC     &7.566&61&4 837&12 249\\
                              &ECP(MRCI)~\cite{dolg-yb2}         &7.544&54&2 662&15 244\\\hline \hline
                                                   
0$_u^{+}(^3\Pi_{u})$          &TZ-ANO-RCC        &8.927&17&  500&16 272\\
                              &QZ-ANO-RCC        &8.575&25&  652&16 236\\
                              &large-ANO-RCC     &8.450&32&  977&16 109\\
                              &ECP(MRCI)~\cite{dolg-yb2}         &8.714&26& 484&18 067\\\hline \hline
                                                   
\hline
\end{tabular}
\caption{\label{tbl:so-caspt2-3p1} SO-CAS(4,8)PT2 adiabatic electronic states of \ce{Yb2} dissociating into the $^1$S+$^3$P$_1$ atomic limit. $\mathrm {r_e}$ denotes the equilibrium bond length, $\mathrm {\omega_e}$ the vibrational frequency, $\mathrm {D_e}$ the potential depth, and $\mathrm {T_e}$ the adiabatic excitation energy, respectively. 
}
\end{table}

\begin{table}[h!]
\centering
\begin{tabular}{|c|c|c|c|c|c|}\hline 
State &basis & $\mathrm {r_e [a_0]}$& $\mathrm {\omega_e [cm^{-1}]}$ & $\mathrm {D_e [cm^{-1}]}$ &$\mathrm {T_e [cm^{-1}]}$ \\ \hline  
                                                   
2$_g(^3\Pi_{g})$              &TZ-ANO-RCC                    &6.964&69&6 303&11 651\\
                              &QZ-ANO-RCC                    &6.909&70&6 685&11 354\\
                              &large-ANO-RCC                 &6.814&72&7 605&10 632\\
                              &ECP(MRCI)~\cite{dolg-yb2}     &6.998&66&5 565&14 437\\\hline \hline
                                                  
0$_u^-(^3\Pi_{u})$            &TZ-ANO-RCC                    &8.626&29&1 403&16 552\\
                              &QZ-ANO-RCC                    &8.528&32&1 549&16 490\\
                              &large-ANO-RCC                 &8.490&32&1 786&16 450\\
                              &ECP(MRCI)~\cite{dolg-yb2}     &8.404&31&1 452&18 551\\\hline \hline
                                                   
1$_u(^3\Pi_{u})$              &TZ-ANO-RCC                    &8.820&23&  902&17 052\\
                              &QZ-ANO-RCC                    &8.668&24&1 012&17 026\\
                              &large-ANO-RCC                 &8.368&30&1 284&16 952\\
                              &ECP(MRCI)~\cite{dolg-yb2}     &8.596&27&1 049&18 954\\\hline \hline

2$_u    (^3\Pi_{u})$          &TZ-ANO-RCC                    &8.931&19&  382&17 572\\
                              &QZ-ANO-RCC                    &8.716&21&  486&17 552\\
                              &large-ANO-RCC                 &8.342&29&  755&17 482\\
                              &ECP(MRCI)~\cite{dolg-yb2}     &8.755&27&  484&19 519\\\hline \hline

1$_g(^1\Pi_{g}, ^3\Sigma_g^+)$ &TZ-ANO-RCC              &12.533&26&   88&17 866\\
                              &QZ-ANO-RCC                   &11.160&18&   81&17 957\\
                              &large-ANO-RCC                &10.943&21&  321&17 916\\
                              &ECP(MRCI)~\cite{dolg-yb2}     &6.837&76&1 129&18 873\\\hline \hline

\hline
\end{tabular}
\caption{\label{tbl:so-caspt2-3p2} SO-CAS(4,8)PT2 adiabatic electronic states of \ce{Yb2} dissociating into the $^1$S+$^3$P$_2$ atomic limit. $\mathrm {r_e}$ denotes the equilibrium bond length, $\mathrm {\omega_e}$ the vibrational frequency, $\mathrm {D_e}$ the potential depth, and $\mathrm {T_e}$ the adiabatic excitation energy, respectively. 
}
\end{table}

\begin{table}[h!]
\begin{threeparttable}
\centering
\begin{tabular}{|c|c|c|c|c|c|}\hline 
State &basis & $\mathrm {r_e [a_0]}$& $\mathrm {\omega_e [cm^{-1}]}$ & $\mathrm {D_e [cm^{-1}]}$ &$\mathrm {T_e [cm^{-1}]}$ \\ \hline  

0$_u^+(^1\Sigma_{u}^+)$       &TZ-ANO-RCC                    &7.173&81&9 089&19 194\\
                              &QZ-ANO-RCC                    &7.378&55&8 765&18 947\\
                              &large-ANO-RCC$^a$             &-&-&-&-\\
                              &ECP(MRCI)~\cite{dolg-yb2}     &7.347&59&7 743&23 713\\\hline \hline
                                                   
1$_u(^1\Pi_{u})$              &TZ-ANO-RCC                    &7.312&66&5 582&22 702\\
                              &QZ-ANO-RCC                    &7.272&67&5 484&22 228\\
                              &large-ANO-RCC                 &7.199&69&6 127&21 328\\
                              &ECP(MRCI)~\cite{dolg-yb2}     &7.170&68&5 001&26 455\\\hline \hline

0$_g^+  (^1\Sigma_{g}^+)$     &TZ-ANO-RCC                    &7.505&65&5 041&23 243\\
                              &QZ-ANO-RCC                    &7.524&57&4 853&22 859\\
                              &large-ANO-RCC                 &7.666&55&5 011&22 444\\
                              &ECP(MRCI)~\cite{dolg-yb2}     &7.514&51&3 549&27 907\\\hline \hline
                                                   
\hline
                                                   
1$_g(^3\Sigma_{g}^+,^1\Pi_g)$ &TZ-ANO-RCC                    &7.852&60&1 917&26 366\\
                              &QZ-ANO-RCC                    &7.802&64&2 200&25 512\\
                              &large-ANO-RCC                 &7.704&69&3 032&24 423\\
                              &ECP(MRCI)~\cite{dolg-yb2}     &8.484&115&9 033&22 422\\\hline \hline
                                                  
\end{tabular}
\caption{\label{tbl:so-caspt2-1p1} SO-CAS(4,8)PT2 adiabatic electronic states of \ce{Yb2} dissociating into the $^1$S+$^1$P$_1$ atomic limit. $\mathrm {r_e}$ denotes the equilibrium bond length, $\mathrm {\omega_e}$ the vibrational frequency, $\mathrm {D_e}$ the potential depth, and $\mathrm {T_e}$ the adiabatic excitation energy, respectively. 
}
 \begin{tablenotes}
      \small
      \item $^a$ not computed due to technical difficulties 
\end{tablenotes}
\end{threeparttable}
\end{table}

\section*{\sffamily \Large CONCLUSIONS AND OUTLOOK}
In this work, we have investigated the electronic structure of atomic and molecular ytterbium using modern, state-of-the-art quantum chemistry methods. 
Our numerical analysis suggests that SO-CASPT2 with inclusion of the 5d orbitals into the active space can accurately reproduce the experimental energy levels of Yb. 
The singlet excited states in the Yb atom can be reliably modeled within the EOM-CCSD approach. 
A similar accuracy in excited state energies and properties can also be obtained with simplified alternatives, such as EOM-pCCD-LCCSD. 

Furthermore, we report a new set of spectroscopic parameters for the ground-state potential energy curve of the Yb$_2$ dimer. 
Our best estimate based on CCSD(T) reference calculations with an uncontracted ANO-RCC basis set gives an optimal bond length of $\mathrm {r_e}=8.814$ bohr, a harmonic vibrational frequencies of $\omega_e=21$ cm$^{-1}$, and a potential energy depth of $\mathrm {D_e}=579$ cm$^{-1}$ and can be considered as the limit of present-day quantum chemistry calculations.  
The CCSD potential energy curve, which results in an elongated bond length and underestimated potential energy depth compared to CCSD(T), can be reliably approximated using the pCCD-LCSSD and CAS(4,8)PT2 approaches. 

The quantum chemical modeling of excited states in the Yb$_2$ dimer remains, however, a remarkable challenge for present-day quantum chemistry. 
First, it is technically challenging to include d-type and f-type orbitals in molecular CASSCF calculations, limiting the manifold of electronic excitations to electron transfer form 6s to 6p atomic orbitals. 
Second, the $^1\Sigma_g^+$ excited state has a double excitation character that is difficult to describe using conventional coupled cluster type methods such as EOM-CCSD.
The EOM-pCCD-LCCSD approach is advantageous here as it provides accurate spectroscopic constants, yet being able to correctly model the doubly excited $^1\Sigma^+$ potential energy curve.
Most importantly, our numerical results indicate that the simplified EOM-pCCD-LCCSD formalism poses an alternative to the conventional EOM-CCSD approach to model excited states.
Specifically, for most excited states, EOM-pCCD-LCCSD (with or without orbital optimization) provides spectroscopic constants that deviate less from the SO-CAS(4,8)PT2 reference values.
This is especially advantageous for excited states with significant bi-excited character, where EOM-CCSD is known to fail.

Finally, we report a new set of reference spectroscopic constants for the low-lying excited states of the Yb$_2$ dimer using the SO-CAS(4,8)PT2 approach. 
Our data is a significant improvement over the existing ECP/SO-MRCI results of Wang and Dolg~\cite{dolg-yb2} as they include an all-electron basis set and a more rigorous treatment of scalar relativistic and electron correlation effects within the CASPT2 approach. 
Moreover, we investigate the convergence of the spectroscopic parameters (optimal bond lengths, vibrational frequencies, potential energy depths, and adiabatic excitation energies) with respect to the size of the basis set, which highlights the need for large basis set when modeling excited state potential energy curves in Yb$_2$. 
We would like to stress that new quantum chemistry methods are desirable that can be used to reliably model the complete set of excited state potential energy curves in challenging molecules like the Yb$_2$ dimer.

High quality potential curves for the Yb$_2$ molecule are critical for future investigations in the fields of cold atomic collisions and ultracold molecules, including an improved description of the strengths and widths of intercombination line optical Feshbach resonances,~\cite{Borkowski2009,Nicholson-2009} searching for routes to Yb$_2$ rovibrational ground state,~\cite{Sage-2009} or calculating the sensitivity of deeply bound molecular clock states to the variation of the proton-to-electron mass ratio~\cite{Zelevinsky2008,Borkowski2018}. Finally, these potential curves provide a valuable starting point for laser-induced fluorescence spectroscopy spectroscopy of ytterbium molecules.
\subsection*{\sffamily \large ACKNOWLEDGMENTS}
P.T.~thanks a POLONEZ 1 research grant (no.~2015/19/P/ST4/02480) financed by Marie-Sk\l{}odowska-Curie COFUND. This project has received funding from the European Union's Horizon 2020 research and innovation programme under the Marie Skłodowska-Curie grant agreement No 665778. K.B.~acknowledges financial support from a Marie-Sk\l{}odowska-Curie Individual Fellowship project no.~702635--PCCDX and a scholarship for outstanding young scientists from the Ministry of Science and Higher Education.
M.B. and P.Sz.Ż.~acknowledge financial support from the National Science Centre, Poland (no.~2017/25/B/ST4/01486). 
The research is part of an ongoing research program of the National Laboratory FAMO in Torun, Poland. 
D.K. acknowledges support from an OPUS grant of the National Science Centre, Poland (no.~DEC-2012/07/B/ST4/01347) and COST Action CM1405 ‘‘Molecules in Motion’’ (MOLIM).

Calculations have been carried out using resources provided by Wroclaw Centre for Networking and Supercomputing (http://wcss.pl), grant nos.~353,~411, and~412.

\end{document}